\documentclass[acus]{JAC2000}

\usepackage{graphicx}

\begin{document}
\title{\flushright{THAP060}\\[15pt] \centering SNS APPLICATION PROGRAMMING 
PLAN\thanks{Work supported by DOE contract DE-AC05-00OR22725}}

\author{C.M. Chu, J. Galambos, J. Wei\thanks{also at BNL}, ORNL, Oak Ridge, TN, 
USA\\
C.K. Allen, P. McGehee, LANL, Los Alamos, NM, USA
}

\maketitle

\begin{abstract}
The architecture for Spallation Neutron Source accelerator physics application 
programs is presented. These high level applications involve processing 
and managing information from the diagnostic instruments, the machine control 
system, models and static databases; they will be used to investigate and control 
beam behavior. Primary components include an SNS global database
and Java-based Application Toolkit, called XAL.
A key element in the SNS application programs is time
synchronization of data used in these applications, due to the short pulse
length (1 ms), pulsed (60 Hz) nature of the device.  The data synchronization 
progress is also presented. 
\end{abstract}

\section{Introduction}
The Spallation Neutron Source (SNS) is a high intensity pulsed accelerator for
neutron production.  To commission and run the SNS efficiently, 
high level physics application software for modeling, integrated operation 
and accelerator physics studies is required; in particular,  
construction of an object-oriented, 
accelerator-hierarchy programming framework. 
Java is chosen as the core programming language because 
it provides object-oriented scope and  existing interfaces to
the controls software ({\it e.g.} Java Channel Access)  and database 
information (JDBC, XML).
The SNS physics application software environment includes
the SNS global database, a Java-based software infrastructure (XAL), and 
existing lattice tools such as Trace-3D and MAD.  
The core part of this environment is the XAL infrastructure, which includes
links to the SNS database, EPICS Channel Access signals, 
shared extensible markup language (XML) files among applications and external 
modeling tools, 
as well as built-in accelerator physics algorithms.  
The present plan for quick on-line modeling during the SNS commissioning is to 
use Trace-3D for the Linac and MAD for the Ring.
Data synchronization at the EPICS level for the SNS pulsed 
nature is also in progress, and will be included in the 
XAL infrastructure later.  

\begin{figure*}[htb]
\centering
\includegraphics*[width=80mm]{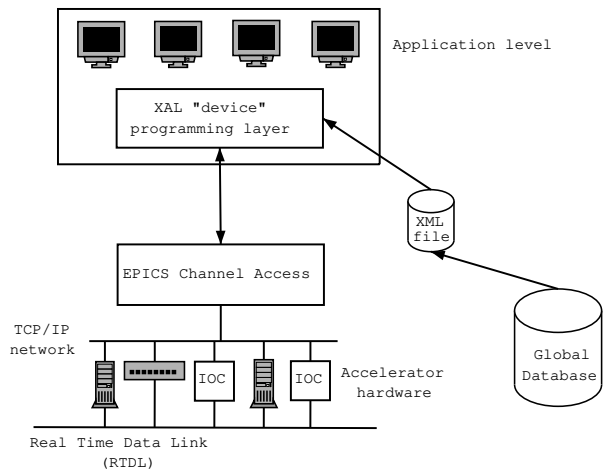}
\caption{Application software infrastructure.} \label{app}
\end{figure*}

\section{SNS Global Database}
The SNS global database contains static information about beam line devices 
(magnets, diagnostics, etc.), power supplies, magnet measurement, global 
coordinates, as well as other accelerator equipment. The table schemas, 
entities and relationships are described in \cite{db}. 
The basic accelerator hierarchy is constructed from the database information. 
For example information for constructing representative beamline sequences, 
their constituent lattice and diagnostic components, and the mapping of 
beamline components to their respective EPICS Process Variables (PVs) all 
comes from the global database.  

Although it is possible to directly query the database from the Java based XAL 
framework, an intermediate XML file containing the accelerator hierarchy 
is created instead. The structure of the XML files is based on 
the XAL class view. The global database to local XML file 
translation is a stand-alone program outside the XAL, which obviates the need 
for each XAL based applications to 
query the database for initialization. 

\section{XAL Infrastructure}
\begin{figure*}[thb]
\centering
\includegraphics*[width=130mm]{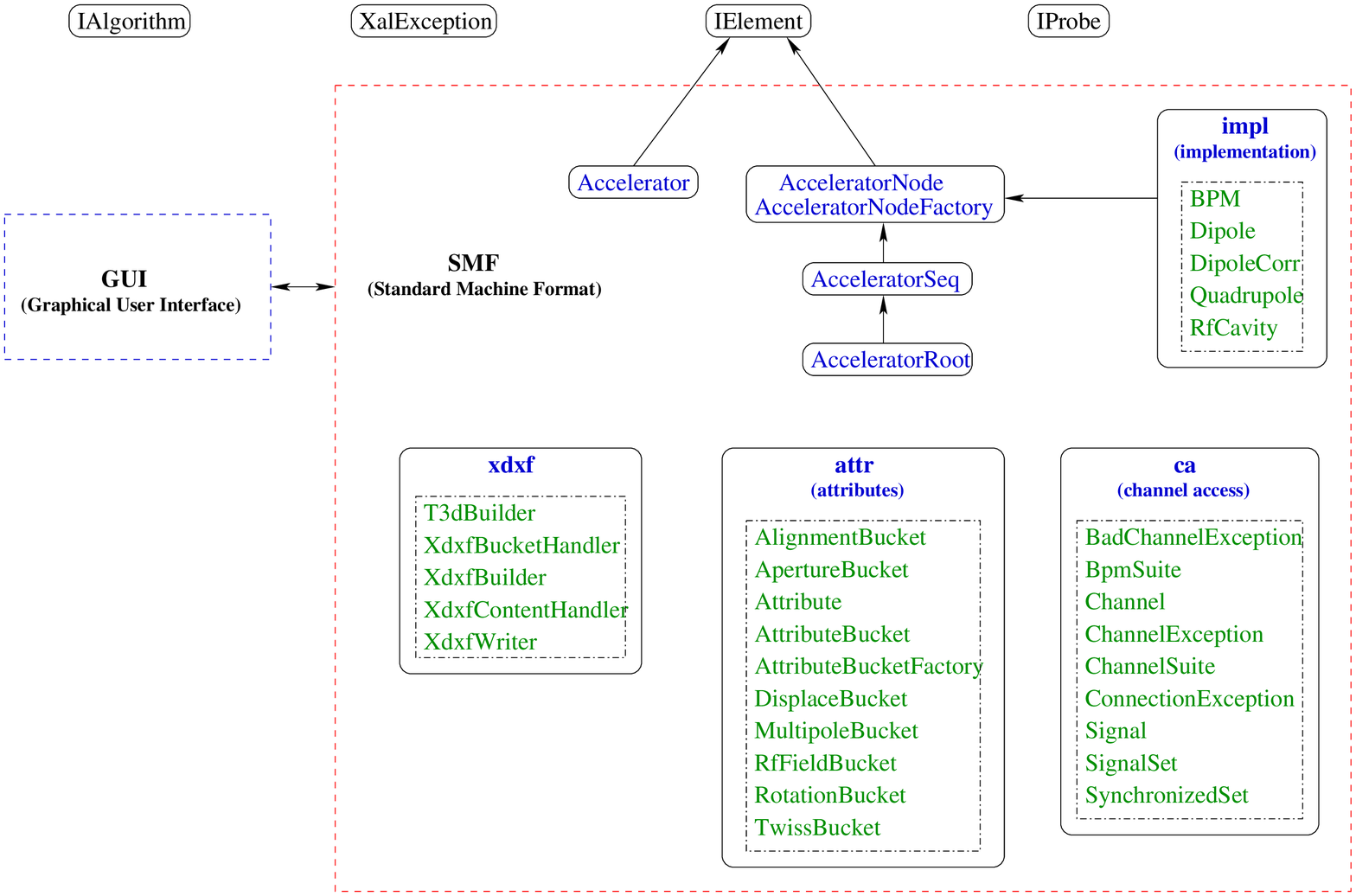}
\caption{XAL infrastructure.} \label{xal_cs}
\end{figure*}

The XAL infrastructure is a Java class structure providing a programming 
interface with an accelerator hierarchy view. XAL is a variant of UAL  
2.0\cite{ual}, 
and detailed API information for the XAL can be found on-line\cite{api}.  
A schematic diagram depicting the XAL infrastructure relationship to other 
accelerator components is shown in Fig.~\ref{xal_cs}.  The XAL provides 
application programs with connections to 
the static data via XML files and the run-time data via Java Channel Access.

The XAL class hierarchy is shown in Fig.~\ref{xal_cs}.
At the top of the XAL class hierarchy 
is the SNS accelerator. The accelerator is composed of different Accelerator 
sequences, {\it e.g.} Medium Energy Beam Transport (MEBT), Drift Tube Linac
(DTL), Ring.  The sequences are composed of nodes, {\it e.g.} 
Quadrupoles, BPMs, Correctors. 
There is a built-in capability to include algorithms in XAL, but initially we 
are using an external model (Trace-3D) for the Linac applications.
Regarding scripting possibilities, XAL class objects directly with Jython
are being tested, without the need for interface code. 

\subsection{EPICS Channel Access}
All the run-time information for the applications will be obtained through 
EPICS Channel Access.  The XAL provides the connectivity to the EPICS Channel 
Access via the channel access (ca) classes as shown in Fig.~\ref{xal_cs}.  
Because the SNS is a pulsed machine, for many applications the
data correlation among pulses is vital.  The ca classes provide both
synchronized and non-synchronized methods for data taking.
The data synchronization will be
described in detail in Section~\ref{sync}.  

\subsection{Links to External Modeling Tools}
Most of the existing accelerator modeling software packages are written in 
languages other than Java.  In order to run applications from Java-based  
XAL, the software packages must be compiled as shared libraries, then 
connected to the shared libraries via the Java Native Interface (JNI). The file
I/O is done through XML parsing provided by XAL, 
for example, storing the calculated result in XML files. 
Thus the information is portable, share-able, and can be accessed 
remotely.  The JNI calls also require arranging the running threads carefully
because programs normally tend to execute its own threads before starting the
non-Java threads.  

\section{Data Synchronization}\label{sync}
\begin{figure*}[htb]
\centering
\includegraphics*[width=60mm]{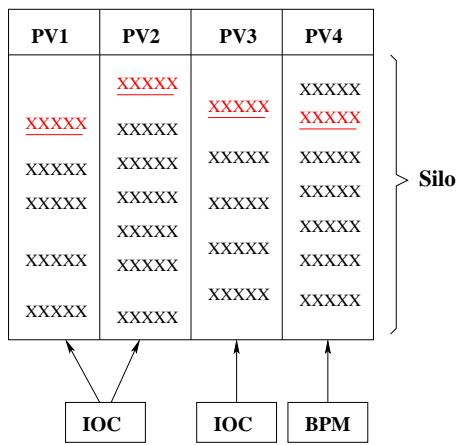}
\caption{Data-silo for data synchronization.} \label{silo}
\end{figure*}

Data synchronization is an important feature for a pulsed accelerator (1 ms 
beam pulses at 60 Hz). The SNS Real Time Data Link will synchronize the clocks 
of all IOCs across the accelerator at 60~Hz rate, 
ensuring a good synchronization of the time-stamps being applied to 
PVs\cite{timing}. However, it may be difficult for high 
level applications to reliably gather sets of data from across the accelerator, 
all from the same pulse. To facilitate this, a data-silo data time correlator 
is being written. The data-silo method is shown schematically in 
Fig.~\ref{silo}.  For a requested PV set, the correlator returns the most 
recent collection of time-correlated data.
    
The behavior of the DataSilo class is configurable
by three parameters: the maximum time to wait since start of request,
maximum width of the time bin, and the maximum number of channels 
allowed to be missing from the synchronized data set.
The correlator is implemented as the C++ DataSilo class which allows
the application's programmer to: 
\begin{itemize}
\item add and remove EPICS process variables from 
the DataSilo set;
\item dynamically define the maximum wait time, maximum bin number, 
and maximum missing bins allowed;
\item obtain the most recent synchronized set (no waiting);
wait up to the maximum time to obtain a synchronized set (blocking)
\item choose the earliest, latest, or mean time stamp from a
synchronized set.
\end{itemize}

\section{Conclusion}
The SNS global database is close to the end of design phase and has been tested
with SNS MEBT data.  The XAL infrastructure is constructed and tested with
a modeling tool, Trace-3D.  The Channel Access part of the XAL will be
tested with simulated IOC signals.  Scripting tools such as Matlab and Python 
will be used in the MEBT commissioning this spring.  

\section{Acknowledgments}
The authors would like to thank 
the SNS Controls and Diagnostics groups for kindly 
providing us all the EPICS, database and other support.
We would also like to thank Dr. N. Malitsky for his help on the early
XAL development.

\end{document}